# Resilient Consensus-Based Target Tracking under False Data Injection Attacks in Multi-Agent Networks

**Amir Ahmad Ghods [1] and Mohammadreza Doostmohammadian [1,] ***

[1] Faculty of Mechanical Engineering, Semnan University, Iran; amir-ghods@semnan.ac.ir (AA Ghods); doost@semnan.ac.ir (M Doostmohammadian)

* Correspondence: doost@semnan.ac.ir

**Abstract:** Distributed target tracking in multi-agent networks plays a critical role in cooperative sensing and autonomous navigation. However, it faces significant challenges in highly dynamic and adversarial setups. This study aims to enhance the resilience of decentralized target tracking algorithms against measurement faults and cyber-physical threats, especially false data injection attacks. We propose a consensus-based estimation algorithm that integrates a nearly-constant-velocity model with saturation-based filtering to suppress impulsive measurement variations and promote robust, distributed state estimation. To counteract adversarial conditions, we incorporate a dynamic false data injection detection and isolation mechanism that uses innovation thresholds to identify and disregard suspicious measurements before they can degrade the global estimate. The effectiveness of the proposed algorithms is demonstrated through a series of simulation-based case studies under both benign and adversarial conditions. The results show that increased network connectivity and higher consensus iteration rates improve estimation accuracy and convergence speed, while properly tuned saturation filters achieve a practical balance between fault suppression and accurate estimation. Furthermore, under localized, coordinated, and transient false data injection attacks, the detection mechanism successfully identifies compromised agents and prevents their data from corrupting the distributed global estimate. Overall, this study illustrates that the proposed algorithm provides a simplified fault-tolerant solution that significantly enhances the accuracy and resilience of distributed target tracking without imposing excessive communication or computational burdens.

## 1. Introduction

In recent years, distributed target tracking has played a crucial role in various applications, including autonomous surveillance, environmental monitoring, and cooperative robotics. These systems rely on distributed estimation and local communication among agents to track moving targets in a scalable, robust, and real-time manner. However, maintaining estimation accuracy and network resilience in distributed systems is significantly challenged by faulty measurement, environmental disturbances, and the increasing risk of cyber-physical threats.

To address these challenges, researchers have examined various architectural strategies for estimation and communication in multi-agent systems. Estimation architectures are generally divided into two primary categories based on how information is shared. The first category is a centralized platform, where a single central node collects data from all agents in the communication network, conducts the necessary computations, and subsequently shares the results with those agents [1-4]. While this approach can achieve high estimation accuracy, it suffers from several critical drawbacks. The reliance on a single node introduces a potential point of failure, requires high computational capacity at that node, and leads to poor scalability and security concerns. As a result, many researchers have shifted their attention toward decentralized frameworks, in which the central node is eliminated and each agent processes data locally while communicating with neighboring agents. This structure improves system scalability and fault tolerance, and significantly reduces the communication overheads. Nonetheless, decentralized systems come with their own set of challenges, including increased network complexity and the need for reliable communication and coordination to ensure accurate and consistent performance across all agents.

Several estimation strategies have been explored to address these challenges. Among the most widely studied are variants of the Kalman filter which are adapted for distributed use, commonly known as Distributed Kalman Filters (DKFs) [5-10]. While these approaches allow each agent to maintain a local estimate, they still face limitations in practice. DKFs typically require high communication bandwidth due to the need for covariance information exchange and are not well-suited to adversarial environments due to their limited robustness and dependence on accurate model assumptions. These factors reduce their scalability and make them vulnerable to both faults and intentional data corruption in dynamic settings.

To address the shortcomings of DKFs, researchers have turned to adaptive and learning-based approaches that aim to improve resilience. For example, model-free adaptive control allows agents to follow reference signals using only local information, making it effective even in the presence of communication delays or denial-of-service attacks [11]. Neural network–based techniques have been employed to approximate unknown nonlinear dynamics in time-varying systems, enabling faster adaptation and more accurate tracking [12]. Likewise, adaptive fuzzy controllers provide a way to estimate unmeasurable states and reduce the impact of cyber-physical attacks, keeping tracking errors within acceptable bounds in large-scale networks [13]. More recently, data-driven and federated learning strategies have been explored to counter more severe disruptions such as Byzantine attacks or actuator faults, often by combining robust aggregation rules with online learning of uncertain system dynamics [14,15]. Despite these advances, such methods often entail trade-offs, including intensive computation, significant communication overhead, or complex agent coordination. These drawbacks limit their practicality in large, resource-constrained networks and leave open the need for simpler, more scalable, and inherently robust alternatives.

To overcome these limitations, researchers have increasingly turned to consensus-based estimation algorithms [16-25]. These approaches use simple local rules and information exchange between neighboring agents to gradually reach a shared understanding across the entire network. Unlike traditional distributed Kalman filters, consensus-based algorithms require less communication, scale more efficiently with larger systems, and are more resilient to issues like unstable connections and noisy measurements. Their simple design makes them particularly suitable for real-world deployments, especially in mobile or resource-constrained multi-agent systems [26].

Despite the advantages offered by decentralized estimation, a significant challenge remains in ensuring the resilience of such systems against cyber-physical threats, particularly false data injection attacks (FDIA). In such situations, an attacker deliberately feeds false data into the system, often by taking control of certain agents or by tampering with their communication channels. When it comes to tracking moving targets, these types of attacks can cause serious errors in estimating position and velocity, which are essential for maintaining accurate and reliable system behavior. As a result, improving the resilience of distributed estimation methods against false data injection has become a key focus in recent research. Numerous studies have explored ways to enhance both the communication and estimation processes within distributed networks. A central approach in this area involves developing mechanisms that can detect and isolate suspicious data inputs quickly, preventing them from affecting the accuracy of the overall network estimation [27-37].

This paper builds upon our previous study in [38], which introduced a decentralized, consensus-based target tracking algorithm employing a nearly-constant-velocity (NCV) model and saturation-based filtering to mitigate impulsive measurement variations. While that earlier work established the baseline framework, it did not incorporate mechanisms to identify or isolate malicious data. In the present study, we extend the framework by introducing a dynamic false data injection (FDI) detection and isolation mechanism that allows the algorithm not only to achieve accurate estimates under nominal conditions but also to remain resilient against adversarial attacks in multi-agent networks. This mechanism enables each agent to evaluate incoming measurements in real time, detect anomalies, and disregard suspicious data before it can corrupt the global estimate. To assess the effectiveness of this extension, we conduct a series of baseline and adversarial case studies, including localized, coordinated, and transient attack scenarios. The results demonstrate substantial improvements compared to [38]; for instance, under widespread coordinated attacks with 50% of agents compromised, the proposed method reduces the mean squared estimation error (MSEE) by about 65% relative to the baseline algorithm. Overall, this paper goes beyond our earlier contribution by providing a more comprehensive solution that improves both the accuracy and the security of distributed target tracking, while keeping communication and computational costs low.

The remainder of this article is organized as follows. Section 2 reviews the system modeling setup, including the dynamic model of the moving target and the derivation of the observation matrix as introduced in our earlier work, along with a brief overview of graph theory and network representation. Section 3 is divided into two subsections. The first subsection outlines the consensus-based estimation algorithm with saturation filtering, as proposed in our previous study, and the second subsection introduces the enhancements made through a false data injection detection and isolation mechanism, including the detection thresholds and agent isolation process. Section 4 presents the simulation setup, baseline and robustness case studies, and the associated results, illustrating the algorithm's tracking accuracy and resilience under different network and attack scenarios, while also summarizing the key findings of the article. Section 5 discusses the resilience, limitations, and potential improvements of the proposed methods, and Section 6 provides the overall conclusion of this work.

## 2. System Modeling

This section introduces the mathematical models used to describe the systems. The target's motion is described using a Nearly-Constant-Velocity (NCV) model, which is a kinematic-based model that incorporates process noise to account for the acceleration uncertainties. The measurement model is based on Time-of-Arrival (TOA) multilateration, enabling agents to estimate distances from the target and build local observation matrices. Finally, we used randomly generated networks to present the multi-agent system. These components together form the foundation for the distributed estimation and fault detection methods which will be presented later on in this study.

### *2.1. Target Dynamics with NCV Model*

The motion of the moving target is modeled using a Nearly-Constant-Velocity framework. This approach employs a kinematic transition model and accounts for system uncertainty by incorporating acceleration terms as process noise inputs. Thus, the unmodeled accelerations are represented as disturbances in the following state-space formulation:

$$x(k+1) = A.x(k) + B.w(k), \quad (1)$$

where $x(k)$ and $x(k+1)$ denote the state vector at the current and next time steps, respectively. $A$ and $B$ are the state transition and input matrices, and $w(k)$ is the process noise vector. Using the NCV model with sampling interval $\Delta$, the matrices $A$ and $B$ are given by:

$$A = \begin{bmatrix} 1 & 0 & 0 & \Delta & 0 & 0 \\ 0 & 1 & 0 & 0 & \Delta & 0 \\ 0 & 0 & 1 & 0 & 0 & \Delta \\ 0 & 0 & 0 & 1 & 0 & 0 \\ 0 & 0 & 0 & 0 & 1 & 0 \\ 0 & 0 & 0 & 0 & 0 & 1 \end{bmatrix} \; and \; B = \begin{bmatrix} \frac{\Delta^2}{2} & 0 & 0 \\ 0 & \frac{\Delta^2}{2} & 0 \\ 0 & 0 & \frac{\Delta^2}{2} \\ \Delta & 0 & 0 \\ 0 & \Delta & 0 \\ 0 & 0 & \Delta \end{bmatrix}, \quad (2)$$

in the next section, we present the measurement model used by the agents to complete the mathematical representation of the system.

### *2.2. Measurement Model Using TOA Multilateration*

In the context of distributed state estimation, each agent independently collects measurement data which is linked to the global state of the system. The process of capturing these local observations can be mathematically described by the following measurement model:

$$y_i(k) = H_i.x(k) + v_i(k), \quad (3)$$

Where $y_i(k)$ denotes the observation vector obtained by agent $i$ at time step $k$. $v_i(k)$, models the measurement noise in the data collected by the agents. $H_i$ is the observation matrix that shows how the system state is projected onto the measurements accessible to agent $i$.

The accurate formulation of the observation matrix $H_i$ is essential to ensure that each agent can accurately interpret and reconstruct the state of the system from its own measurements. In this study, we construct $H_i$ using a multilateration approach based on the Time-of-Arrival (TOA) principle, as introduced in [39]. In this setup, the target continuously sends out a signal that moves through the environment at a known speed. When agents detect this signal, they log the time at which it arrives, enabling the calculation of their respective distances from the source. These distance estimates form the basis for both localization and the mathematical formulation of the measurement model by capturing the differences in signal arrival times at various agents.

This derivation is visually illustrated in Figure 1, which shows a scenario where three agents work together to determine the location of a moving target. The figure demonstrates how sending and receiving the beacon signal helps the agents understand their positions relative to the target.

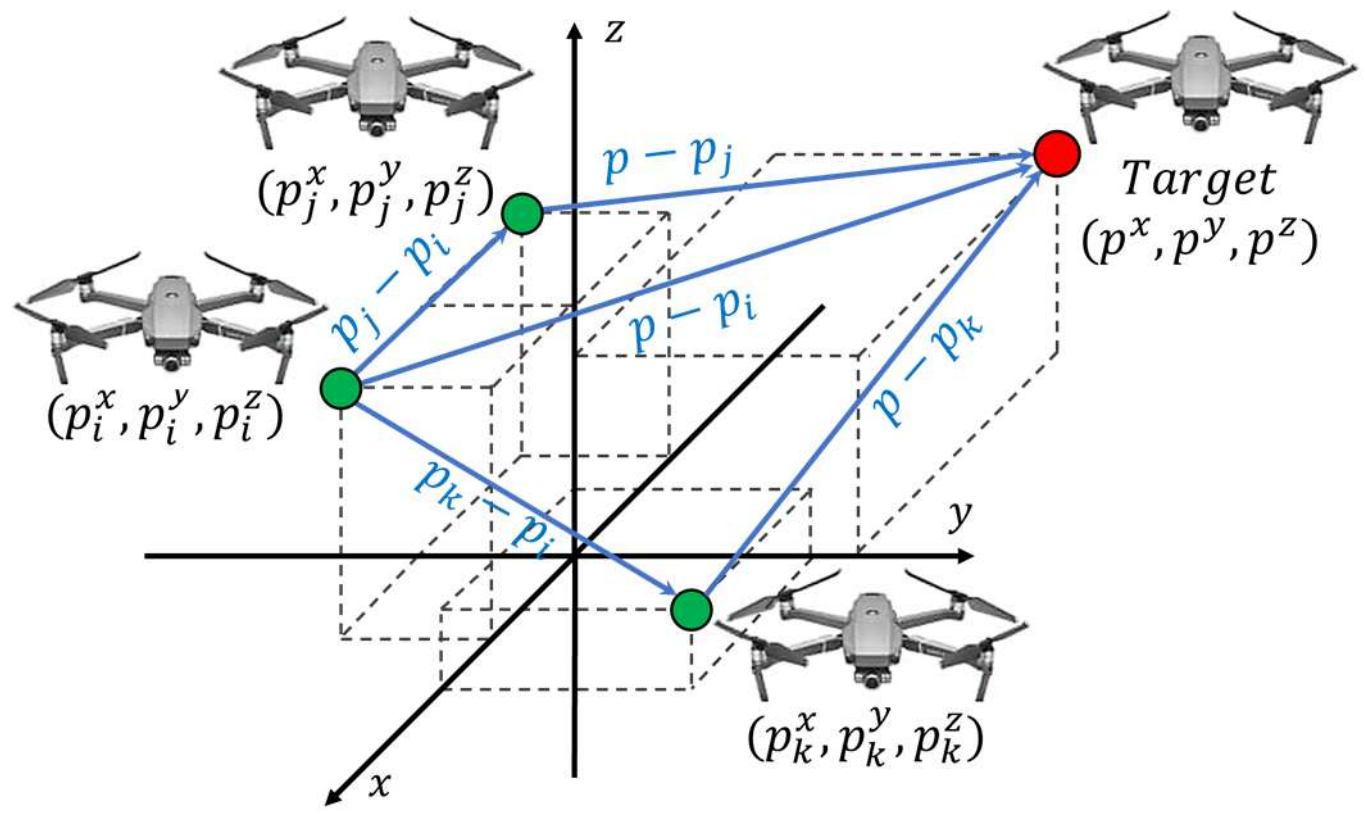

**Figure 1.** Illustration of the multilateration approach used to construct the observation matrix.

For practical implementation, we assume that each agent exchanges data only with its set of immediate neighbors, reflecting a realistic scenario in localized and scalable multi-agent networks. This neighbor-limited information flow allows for the observation matrix to be defined in a linearized format as below:

$$\begin{cases} p_{j,i}^x = p_j^x - p_i^x \\ p_{j,i}^y = p_j^y - p_i^y, \\ p_{j,i}^z = p_j^z - p_i^z \end{cases} \quad and\ H_i = \begin{bmatrix} p_{j_1,i}^T & 0_3^T \\ \vdots & \vdots \\ p_{j_{|N_i|},i}^T & 0_3^T \end{bmatrix}, \tag{4}$$

where $p_{n,i} = (p_{n,i}^x, p_{n,i}^y, p_{n,i}^z)$ expresses the differences of positions between neighbor $n$ and agent $i$. The notation $|N_i|$ denotes the immediate neighboring set for the agent $i$. This setup ensures that each agent's observation matrix is customized to its local surroundings, using only data that can be directly measured or shared within a distributed system.

*2.3. Graph Theory and Network Representation*

In this section, we provide an overview of graph theory and describe the specific network configuration employed in this study. Multi-agent systems are modeled within the framework of graph theory as $\mathcal{G} = \{\mathcal{N},\ \mathcal{L},\ \mathcal{W}\}$, where $\mathcal{N} = \{1, \dots, n\}$ denotes the set of nodes, $\mathcal{L} = \{(i,j) \mid i \to j; i,j \in \mathcal{N}\}$ represents the set of links (edges) among nodes, and $\mathcal{W} = [w_{ij}]$ corresponds to the weight matrix quantifying relevant attributes of the connections, such as distance.

For the purposes of this work, all network topologies are generated using the Erdős–Rényi random graph model. In This model $G(n,p)$, a graph is formed from a fixed set of nodes $n$, where each possible edge between node pairs is included independently with probability $p$. This approach offers analytical simplicity, scalability, and serves as a standard benchmark for examining the effects of random connectivity and network structure in multi-agent systems.

In the following section, we provide a detailed demonstration of the estimation and fault detection algorithms.

## 3. Proposed Estimation and Fault Detection Algorithms

In this section, we first introduce the Consensus-Based Estimation with Saturation Filtering, which was comprehensively explained in our previous work in [38]. We then enhance the performance of this method by incorporating a False Data Injection Detection mechanism. This addition aims to improve the robustness of our approach in adversarial environments and against cyber-physical threats, particularly False Data Injection Attacks on agents, which typically appear as sharp anomalies in the innovation during the tracking of a moving target. For clarity and ease of reference, Table 1 summarizes all major variables, parameters, and symbols used throughout the manuscript.

**Table 1.** Summary of key parameters and variables used throughout the manuscript.

| Symbol | Description |
|---|---|
| $x(k)$ | State vector of the target at time step $k$ |
| $\hat{x}_i(k)$ | Estimated state of the target by agent $i$ at time step $k$ |
| $\lvert N_i \rvert$ | Set of immediate neighbors of agent $i$ |
| $A$ | State transition matrix |
| $B$ | Input matrix |
| $H_i$ | Observation matrix for agent $i$ |
| $y_i(k)$ | Measurement/Observation vector of agent $i$ at time step $k$ |
| $v_i(k)$ | Measurement noise of agent $i$ |
| $w(k)$ | Process noise |
| $g_i(k)$ | Saturation gain agent $i$ at time step $k$ |
| $\xi$ | Observation confidence parameter |
| $L$ | Number of consensus iterations |
| $\varepsilon$ | Consensus step size |
| $\gamma_i(k)$ | Innovation of agent $i$ at time step $k$ |
| $\varphi_i(k)$ | Dynamic detection threshold for agent $i$ at time step $k$ |
| $b_w$ | Upper bounds on the process noise |
| $b_v$ | Upper bounds on the measurement noise |
| $f(k)$ | Global consensus mismatch at time step $k$ |
| $\bar{\rho}_{k,i}$ | Estimation error bound of agent $i$ at time step $k$ |
| $\delta_{k,i}$ | Local consensus mismatch of agent $i$ at time step $k$ |
| $e_i$ | Mean Squared Estimation Error (MSEE) of agent $i$ |
| $e_{Avg}$ | Average Mean Squared Estimation Error of all agents |

*3.1. Consensus-Based Estimation with Saturation Filtering*

To provide context, we briefly summarize the distributed filtering framework in our earlier work. This framework operates in two steps: updating the state estimate using local observations (observation update) and then refining estimates by exchanging information between neighboring agents (estimation consensus).

During the observation update, each agent integrates its own measurement $y_i(k)$ through a saturation-based filtering for the purpose of limiting the effect of corrupted data. The update rule for agent $i$ at time step $k$ is computed as:

$$\hat{x}_i(k) = A\hat{x}_i(k-1) + g_i(k)H_i^T\big(y_i(k) - H_iA\hat{x}_i(k-1)\big), \tag{5}$$

where $\hat{x}_i(k)$ represents the estimated state, $A$ is the known state transition matrix obtained from equation (2), $H_i$ is the observation matrix for agent $i$ obtained from equation (4), and $g_i(k)$ is a saturation gain that addresses the innovation magnitude which is defined by the rule:

$$g_i(k) = \begin{cases} 1, if\ \lvert y_i(k) - H_iA\hat{x}_i(k-1)\rvert \le \xi \\ \dfrac{\xi}{\lvert y_i(k) - H_iA\hat{x}_i(k-1)\rvert}, otherwise. \end{cases} \tag{6}$$

in this format, $\xi > 0$ is the observation confidence parameter. This parameter restricts the influence of large innovations, which may be caused by measurement noise or adversarial interference, thereby enhancing the filter's robustness against such disturbances.

Following the observation update, agents engage in an iterative consensus process to improve estimation accuracy by exchanging information with neighboring agents. This consensus step is performed multiple times within a single time iteration $k$. At consensus iteration $l$, agent $i$'s estimate is updated by combining its own estimate with those of its immediate neighbors $N_i$:

$$\hat{x}_{i,l}(k) = \hat{x}_{i,l-1}(k) - \varepsilon \sum_{J \in N_i} \left( \hat{x}_{i,l-1}(k) - \hat{x}_{J,l-1}(k) \right), \quad (7)$$

in this expression, $\hat{x}_{i,l}(k)$ is the updated estimate, $\varepsilon$ is a small positive which controls the step size of the consensus update, and the summation adjusts the estimate toward the average of neighbors' estimates. Performing multiple consensus steps promotes agreement across the network and improves accuracy compared to using local information alone.

By integrating the saturation-based observation filter with the consensus-based estimation step, the proposed method achieves robustness against corrupted measurements while effectively fusing information across agents. This combined approach forms the foundation of our decentralized mobile target tracking algorithm based on the NCV model. The detailed steps of this method are presented in Algorithm (1).

**Algorithm 1.** Decentralized mobile target tracking using the Consensus-Based Estimation Filter algorithm.

**Initialize variables** = $(\hat{x}_i(0),\ \xi,\ L,\ \varepsilon)$
**for** $k = 1, 2, \ldots$ **do**
  **for** $i = 1, \ldots, N$ **do**
    **Observation Matrix Calculation:**
     **for** $j = 1, \ldots, |N_i|$ **do**
     Compute $\boldsymbol{H_i}$ **"**from Equation (4)"
     **End**
    **Measurement Update with Saturation Filtering:**
    $g_i(k) = min\left(1, \frac{\xi}{|y_i(k) - H_i A \hat{x}_i(k-1)|}\right)$
    $\hat{x}_i(k) = A\hat{x}_i(k-1) + g_i(k) H_i^T \left( y_i(k) - H_i A \hat{x}_i(k-1) \right)$
    **Estimate Consensus:**
     Let $\hat{x}_{i,0}(k) = \hat{x}_i(k)$
     **for** $l = 1, \ldots, L$ **do**
     $\hat{x}_{i,l}(k) = \hat{x}_{i,l-1}(k) - \varepsilon \sum_{J \in N_i} \left( \hat{x}_{i,l-1}(k) - \hat{x}_{J,l-1}(k) \right)$
     **End**
     Let $\hat{x}_i(k) = \hat{x}_{i,L}(k)$
  **End**
**End**

*3.2. False Data Injection Detection and Isolation*

To improve the resilience of Algorithm (1) against cyber-physical threats, particularly false data injection attacks (FDIA), we introduce a detection and isolation mechanism. The core idea is to verify whether each agent's local measurement is consistent with its predicted observation before it is incorporated into the distributed estimation process. If a measurement deviates significantly from what the model expects, it is considered potentially compromised and excluded from the update step.

At each time step $k$, agent $i$ computes an innovation value, which represents the difference between the current measurement and the predicted output based on the agent's previous state estimate. This innovation is given by:

$$\gamma_i(k) = y_i(k) - H_i A \hat{x}_i(k-1), \quad (8)$$

where $y_i(k)$ is the measurement received by agent $i$, $H_i$ is the observation matrix, $A$ is the state transition matrix defined by the Nearly-Constant-velocity model, and $\hat{x}_i(k-1)$ is the predicted state estimate from the previous step.

To assess whether the innovation $\gamma_i(k)$ is acceptable, it is compared against a dynamic detection threshold $\varphi_i(k)$. The threshold is defined as:

$$\varphi_i(k) = \left||A|\right|.(\bar{\rho}_{k-1,i} + f(k-1) + b_w + b_v), \tag{9}$$

where $\left||A|\right|$ is the norm of the state transition matrix $A$. The term $\bar{\rho}_{k-1,i}$ denotes the estimation error bound at the previous time step for agent $i$, and $f(k-1)$ represents the consensus mismatch across the network, which measures how far apart the agents' estimates are at time $k-1$. The constants $b_w$ and $b_v$ represent predefined upper bounds on the process and measurement noise, respectively.

The consensus mismatch term $f(k-1)$ is computed as the maximum deviation between any two agents' state estimates at the previous time step and is calculated as:

$$f(k-1) = \max_{i,j\in N} \left||\hat{x}_i(k-1) - \hat{x}_j(k-1)|\right|, \tag{10}$$

where $N$ denotes the set of all agents in the network. This term reflects the global estimation disagreement and captures the worst-case divergence among the agents.
The estimation error bound $\bar{\rho}_{k,i}$ is updated recursively at each time step using the following relation:

$$\bar{\rho}_{k,i} = \left||A|\right|.\bar{\rho}_{k-1,i} + \varepsilon.\delta_{k-1,i} + b_w, \tag{11}$$

here, $\varepsilon$ is the consensus step size and $\delta_{k-1,i}$ is the local consensus disagreement for agent $i$, representing the maximum deviation between its own estimate and those of its immediate neighbors of $N_i$. This local mismatch is calculated as:

$$\delta_{k-1,i} = \max_{i,j\in N_i} \left||\hat{x}_i(k-1) - \hat{x}_j(k-1)|\right|, \tag{12}$$

if the magnitude of the innovation exceeds the threshold $|\gamma_i(k)| > \varphi_i(k)$), the agent is flagged as compromised. In this case, the agent does not incorporate its current measurement into the state update. Instead, it uses the best prediction of the dynamic model:

$$\hat{x}_i(k) = A\hat{x}_i(k-1), \tag{13}$$

this action effectively isolates the compromised measurement, preventing it from corrupting the overall consensus process.

If the innovation falls within the acceptable threshold, the agent continues with the normal measurement update and consensus-based fusion steps as outlined in the original Algorithm (1). By applying this detection and isolation process at every iteration, the estimation process becomes capable of automatically identifying and isolating faulty measurements. The complete procedure, incorporating this fault-resilient mechanism, is summarized in Algorithm (2).

In the next section, we take a closer look at how Algorithms (1) and (2) perform in a range of testing scenarios. These fall into two main categories: tests under normal conditions with varying network setups, and tests where agents are subjected to false data injection attacks. This approach allows us to evaluate how effectively the algorithms handle disruptions, adapt to changing conditions, and detect faulty information. The analysis aims to assess estimation accuracy, convergence behavior, and overall resilience in a decentralized framework.

**Algorithm 2.** Decentralized Mobile Target Tracking Using Consensus-Based Estimation Algorithm with Fault Detection.

**Initialize variables =** $(\hat{x}_i(0),\ \xi,\ L,\ \varepsilon)$
**for** $k = 1, 2, \ldots$ **do**
 **for** $i = 1, \ldots, N$ **do**
  **Observation Matrix Calculation:**
  **for** $j = 1, \ldots, |N_i|$ **do**
  Compute $\boldsymbol{H_i}$ "from Equation (4)"
  **End**
  **Innovation and Fault Detection:**
  Compute $\gamma_i(k)$ and $\varphi_i(k)$ "from Equations (8) and (9)"
  **if** $|\gamma_i(k)| > \varphi_i(k)$ **then**
  $\hat{x}_i(k) = A\hat{x}_i(k-1)$
  **Else**
  **Measurement Update with Saturation Filtering:**
  $g_i(k) = min\left(1, \frac{\xi}{|y_i(k) - H_i A\hat{x}_i(k-1)|}\right)$
  $\hat{x}_i(k) = A\hat{x}_i(k-1) + g_i(k) H_i^T \left(y_i(k) - H_i A\hat{x}_i(k-1)\right)$
  **Estimate Consensus:**
  Let $\hat{x}_{i,0}(k) = \hat{x}_i(k)$
  **for** $l = 1, \ldots, L$
  $\hat{x}_{i,l}(k) = \hat{x}_{i,l-1}(k) - \varepsilon \sum_{J \in N_i} \left(\hat{x}_{i,l-1}(k) - \hat{x}_{J,l-1}(k)\right)$
  **end if**
  Let $\hat{x}_i(k) = \hat{x}_{i,L}(k)$
 **End**
**End**

## 4. Case Studies and Results

This section presents a comprehensive evaluation of the proposed decentralized, consensus-based target tracking algorithm under a range of operating conditions. The evaluation is divided into two main categories. The first category, referred to as the Baseline Case Studies, investigates the performance of Algorithm (1) in attack-free environments. This part examines the effects of key design parameters, such as graph connectivity, the number of consensus iterations, and the saturation filter settings, on the accuracy of target tracking and the convergence behavior of the distributed estimates.

The second category, referred to as the Robustness Case Studies, evaluates the performance of both Algorithms (1) and (2) under various false data injection attacks. This part emphasizes the resilience and adaptability of the proposed framework when facing cyber-physical threats and highlights its effectiveness in mitigating the impact of faulty measurements.

Performance is assessed using the Mean Squared Estimation Error (MSEE), defined in Equation (14), the Average Mean Squared Estimation Error given in Equation (15), and the innovation norm as defined in Equation (8):

$$e_i = \frac{1}{k} \sum_{k=1,2,\ldots} \left(\hat{x}_i(k) - x_i(k)\right)^2, \tag{14}$$

$$e_{Avg} = \frac{1}{|N|} \sum_{i=1}^{|N|} e_i\,, \tag{15}$$

all simulations were implemented and executed in the MATLAB environment. The simulation setup reflects realistic network conditions and includes multiple agent configurations. Finally, a Summary and Discussion of the Results is provided to show clear comparative analysis.

*4.1. Baseline Case Studies (1–3): Without FDIA*

This set of experiments evaluates the performance of Algorithm (1) under benign conditions, where no false data injection attacks occur. The purpose of these baseline studies is to examine the influence of key design parameters, including the communication topology, the number of consensus iterations, and the configuration of the saturation filter on the estimation accuracy and convergence behavior. These tests provide a foundation for understanding the algorithm's behavior under ideal conditions, before introducing adversarial scenarios.

4.1.1. Case Study 1 – Effect of Graph Connectivity

This case study investigates how the density of the communication graph affects the tracking performance. Two random graphs with 20 agents are generated: one with a connection probability of $p = 0.7$, representing a highly connected network (Fig. 2-b), and one with a low connectivity of $p = 0.3$ (Fig. 2-a). The corresponding average MSEE results for these configurations are shown in Fig. 2-c and Fig. 2-d.

The results indicate that a higher level of connectivity improves estimation accuracy and accelerates convergence by allowing more information to be exchanged between agents. However, this enhanced performance requires greater communication bandwidth and computational resources. Therefore, designers must carefully balance the trade-off between estimation performance and the communication costs when choosing the network topology.

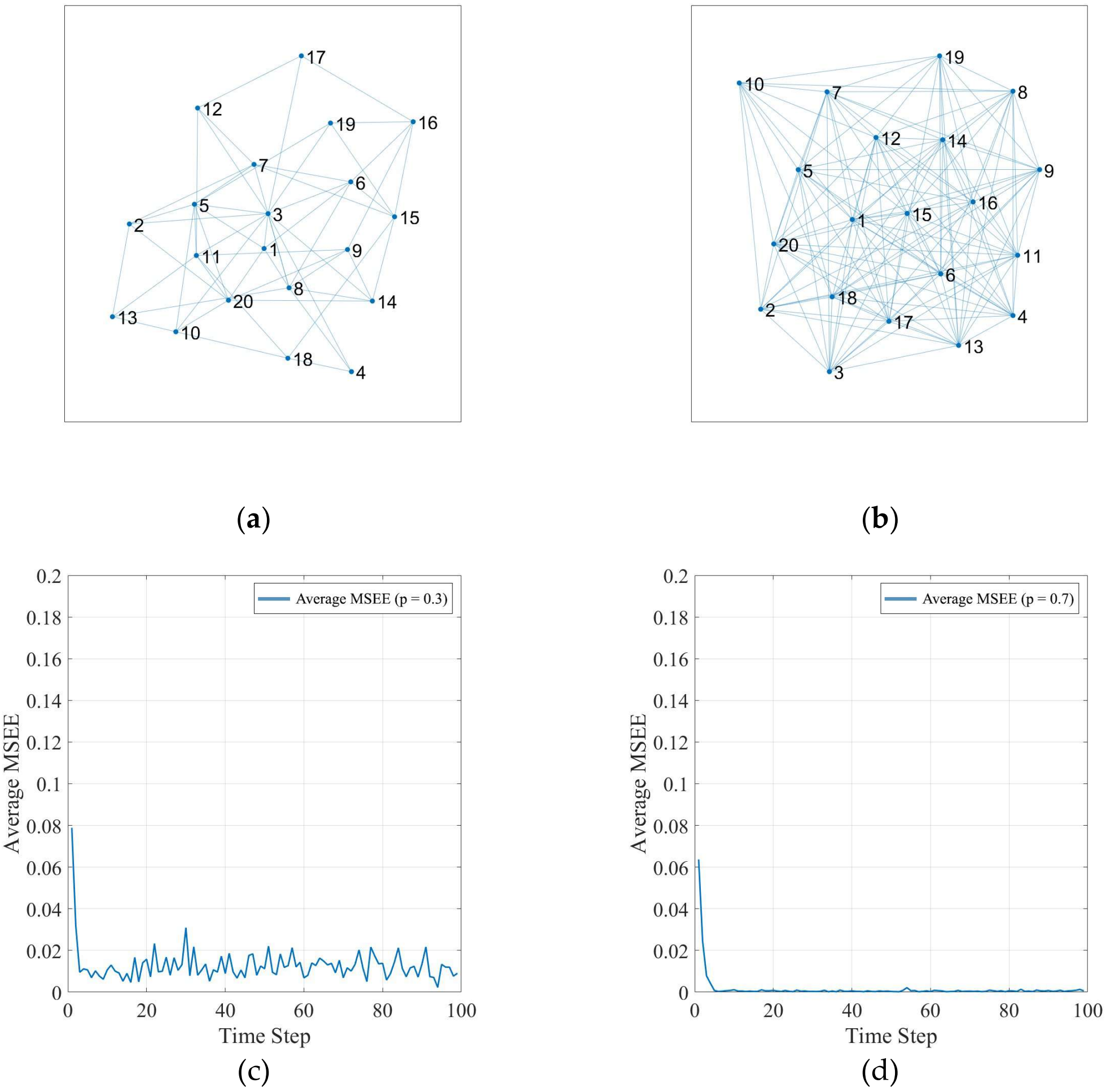

**Figure 2.** Effect of network connectivity on tracking accuracy: (a) Graph with $p = 0.3$; (b) Graph with $p = 0.7$; (c) Average MSEE for the low-density network (shown in blue lines); (d) Average MSEE for the high-density network (shown in blue lines).

4.1.2. Case Study 2 – Number of Consensus Iterations

This experiment examines the effect of the number of consensus iterations, denoted by $L$, which controls how often each agent exchanges information with its immediate neighbors between successive time steps. Two different settings are tested: a lower communication rate ($L = 10$) and a higher rate ($L = 100$).

The results for a random communication network in Fig. 3-a, indicate that using fewer iterations slows the convergence process and yields higher estimation errors (Fig. 3-b and Fig. 3-c), whereas increasing the number of iterations improves the convergence rate and reduces estimation errors (Fig. 3-d and Fig. 3-e). However, this performance gain requires increased communication and computation. Therefore, the number of iterations must balance accuracy with system constraints.

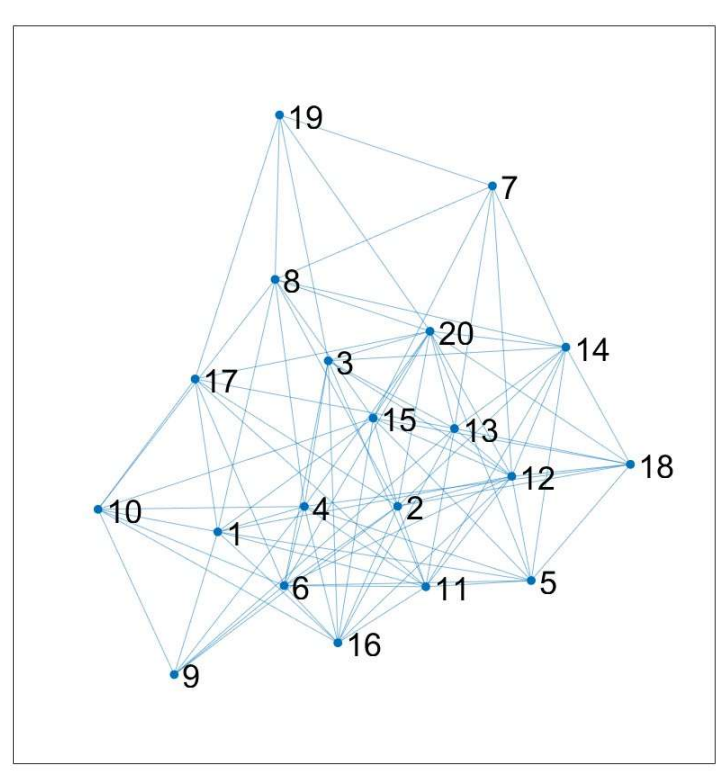

(a)

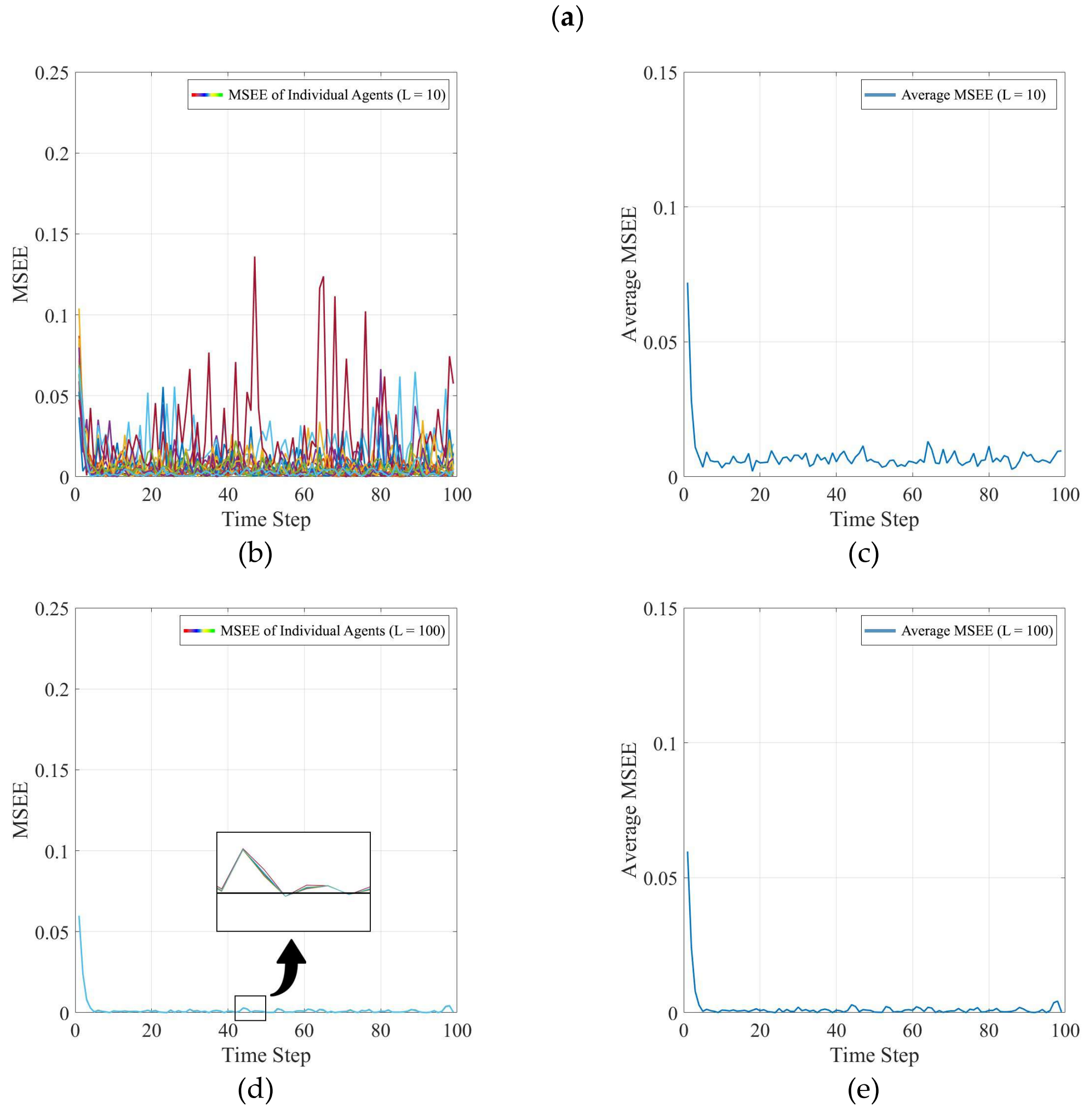

(b) (c)

(d) (e)

**Figure 3.** Impact of the number of consensus iterations on estimation performance: (a) Communication graph; (b–c) MSEE of individual agents and the average MSEE of all agents for $L = 10$, shown in colored and blue curves, respectively; (d–e) MSEE of individual agents and the average MSEE of all agents for $L = 100$, shown in colored and blue curves, respectively.

#### 4.1.3. Case Study 3 – Saturation Filter Parameters

In this final baseline case, the influence of the saturation filter's configuration on tracking robustness and sensitivity is investigated. Two filter settings are tested: a permissive (soft) threshold with ($\xi = 4$) and a conservative (harsh) threshold with ($\xi = 2$). These settings control how much deviation is tolerated in received data before it is attenuated or discarded.

The results are shown for a random communication network in Fig. 4-a. Under the soft threshold setting ($\xi = 4$), the system achieves higher estimation errors, as seen in the innovation norm and average MSEE plots (Fig. 4-b and Fig. 4-c), and it is more vulnerable to faulty measurements. By contrast, the harsh threshold setting ($\xi = 2$) yields smoother innovation norms and more stable behavior (Fig. 4-d and Fig. 4-e), at the cost of less sensitivity to measurement innovations. This case highlights the critical trade-off between filter sensitivity and robustness.

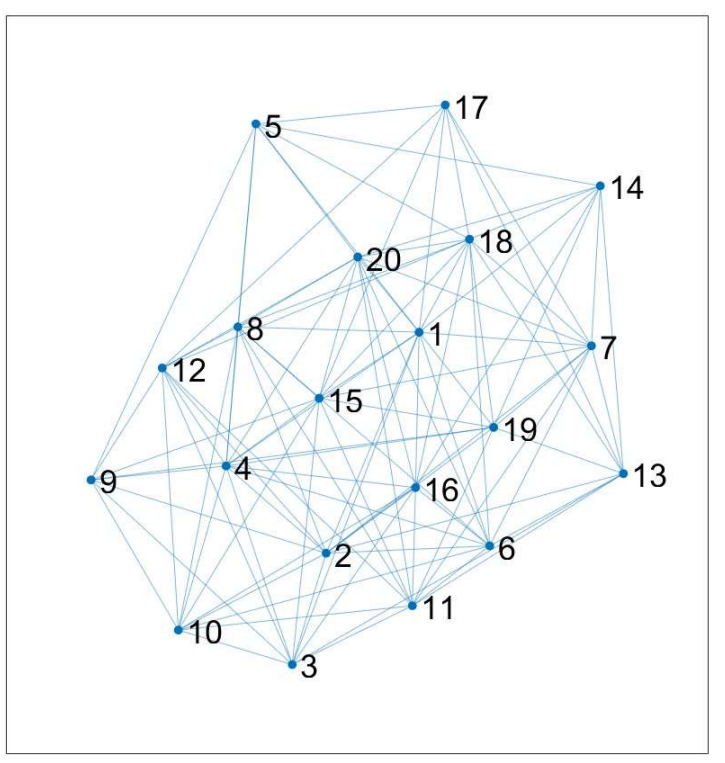

(**a**)

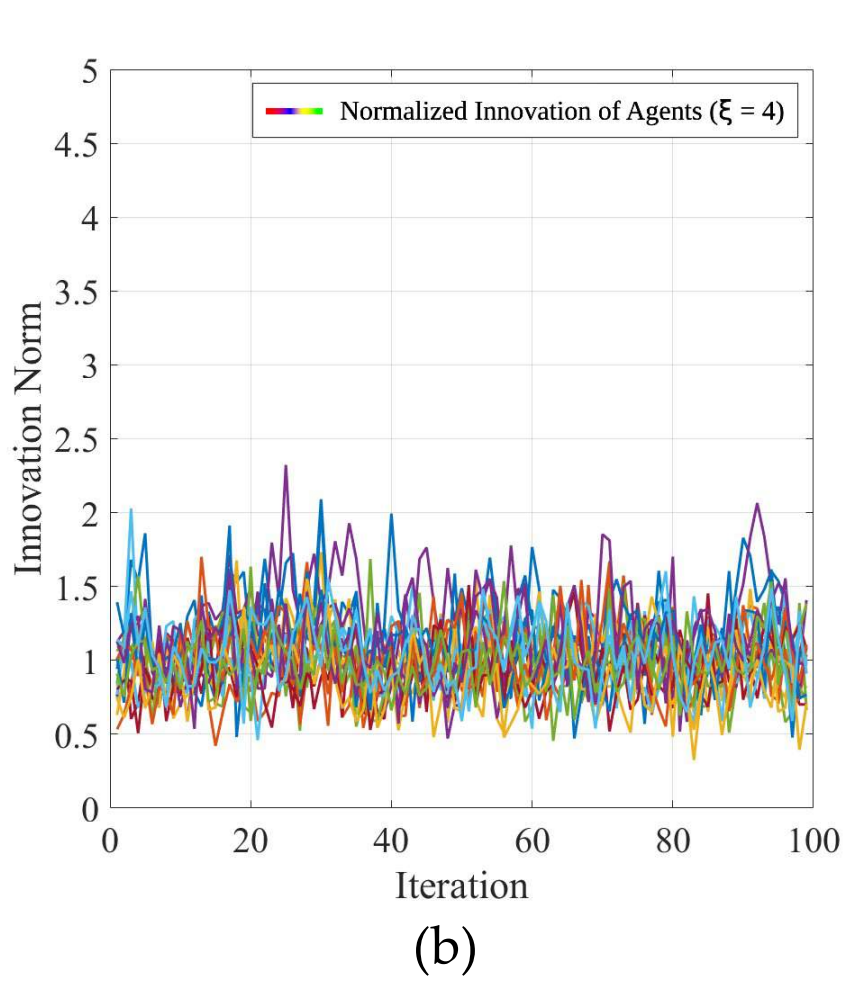

(b)

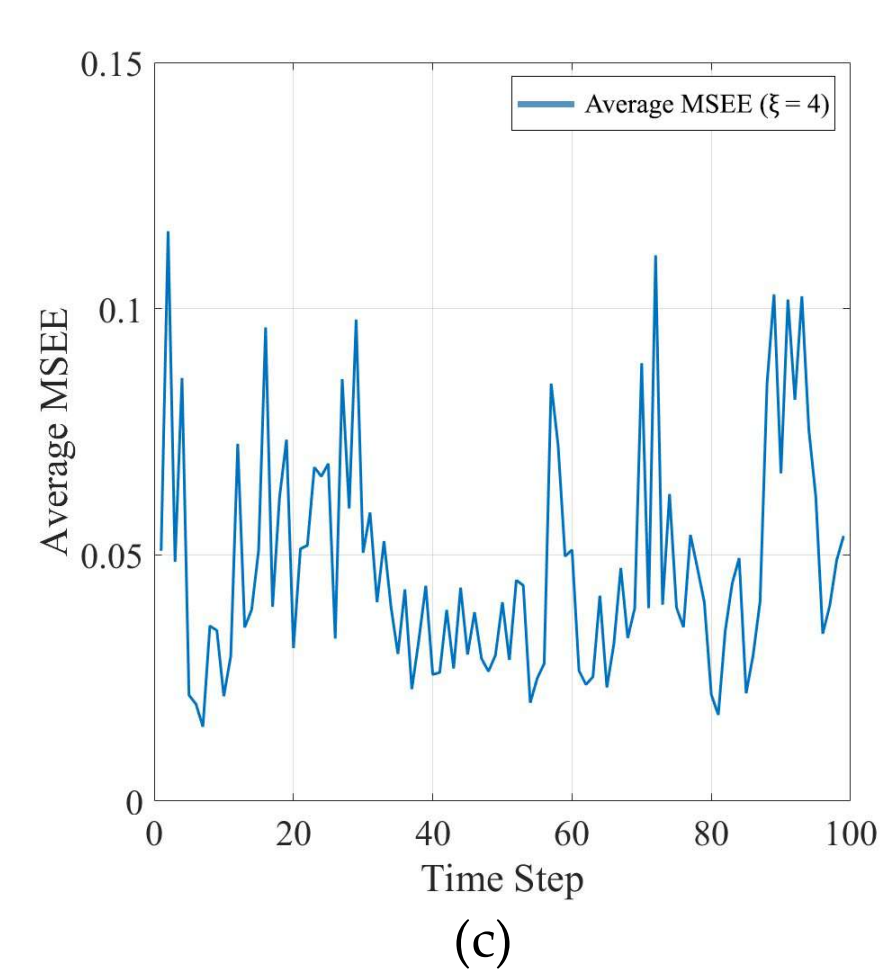

(c)

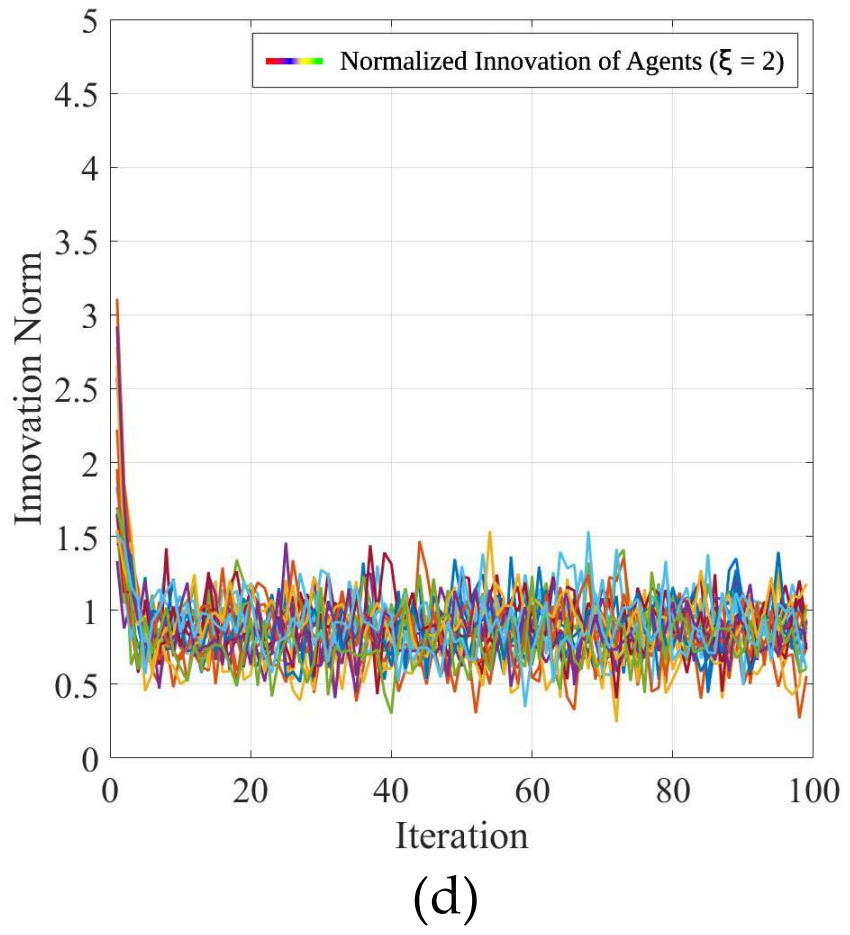

(d)

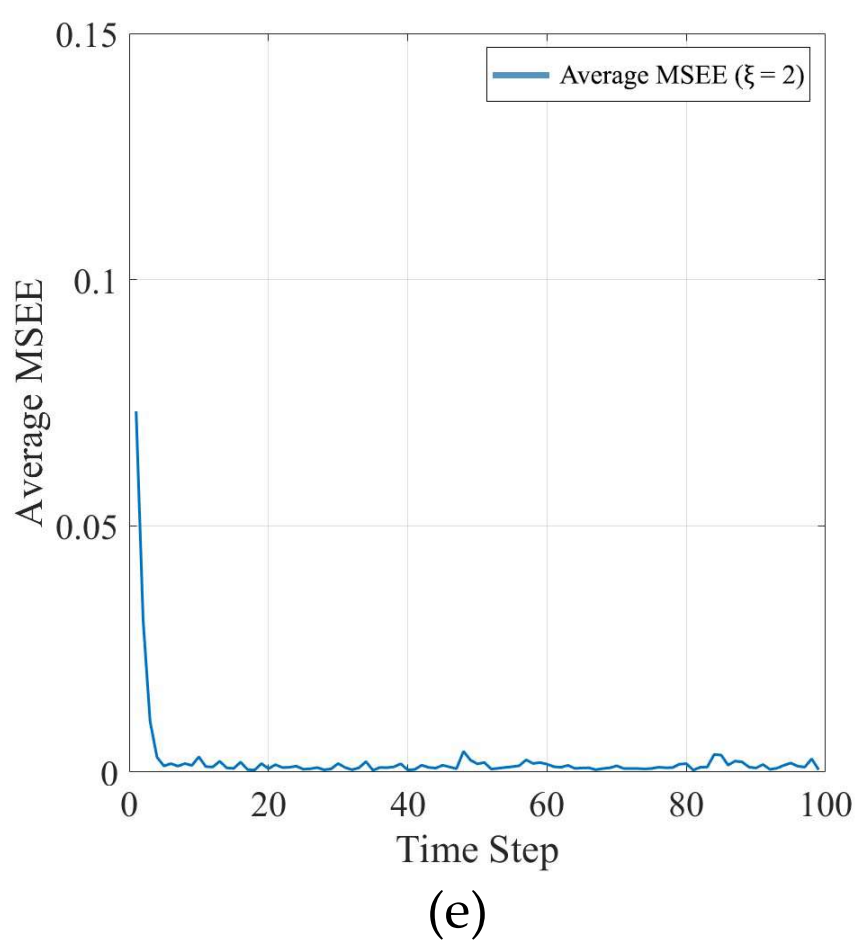

(e)

**Figure 4.** Effect of saturation filter parameters on tracking performance: (a) Network graph; (b–c) Normalized innovation of individual agents and average MSEE of all agents for the soft filter ($\xi = 4$), shown in colored and blue curves, respectively; (d–e) Normalized innovation of individual agents and average MSEE of all agents for the harsh filter ($\xi = 2$), shown in colored and blue curves, respectively.

### *4.2. Robustness Case Studies (4–6): With FDIA*

In this section, we extend our experiments to evaluate the performance of the Algorithm (2) alongside the baseline Algorithm (1), under adversarial conditions. Our focus is on cyber-physical threats posed by false data injection attacks from an external attacker. FDIA can cause substantial disturbances in measurements, creating anomalies or sharp spikes in the data that degrade the accuracy of the state estimates.

In the first case, we introduce a localized false data injection attack of high magnitude to show how the detection and isolation mechanism protects the estimates and improves the reliability of the communication system under cyber-physical threats. The second case considers a widespread coordinated attack where half of the agents receive false data at varying magnitudes, allowing us to assess the algorithm's resilience when a significant portion of the network is compromised. Finally, the third case investigates the algorithm's behavior under transient attacks that last for only a short period of time. In each case, we evaluate and compare the performance of both algorithms to highlight the effectiveness of the proposed detection mechanism under different adversarial scenarios.

#### 4.2.1. Case Study 4 – Localized False Data Injection

In this case study, we assume that two randomly selected agents are subjected to a high-magnitude false data injection attack after a specified time step ($k > 40$). The objective is to demonstrate how the detection and isolation mechanism can identify and isolate the compromised agents so that their corrupted data does not propagate through the network or degrade the overall consensus-based state estimation.

The detection process on a randomly generated network is illustrated in Fig. 5-a. The innovation norm for each agent is compared against the dynamic threshold explained in Equation (9), and any value exceeding this threshold is flagged as compromised, as shown in Fig. 5-b. Figures 5-c and 5-d present the MSEE over time for the algorithms without and with FDIA detection, respectively. The results clearly show that the proposed detection and isolation method effectively mitigates the impact of the attacks, achieving lower errors than the baseline algorithm without detection.

Figure 5-e compares the average MSEE across all agents for both algorithms, further confirming the significant performance improvements provided by the detection

mechanism under cyber-physical threats. Finally, Fig. 5-f illustrates the detection heat map, which correctly identifies agents 5 and 17 as under attack after $k > 40$.

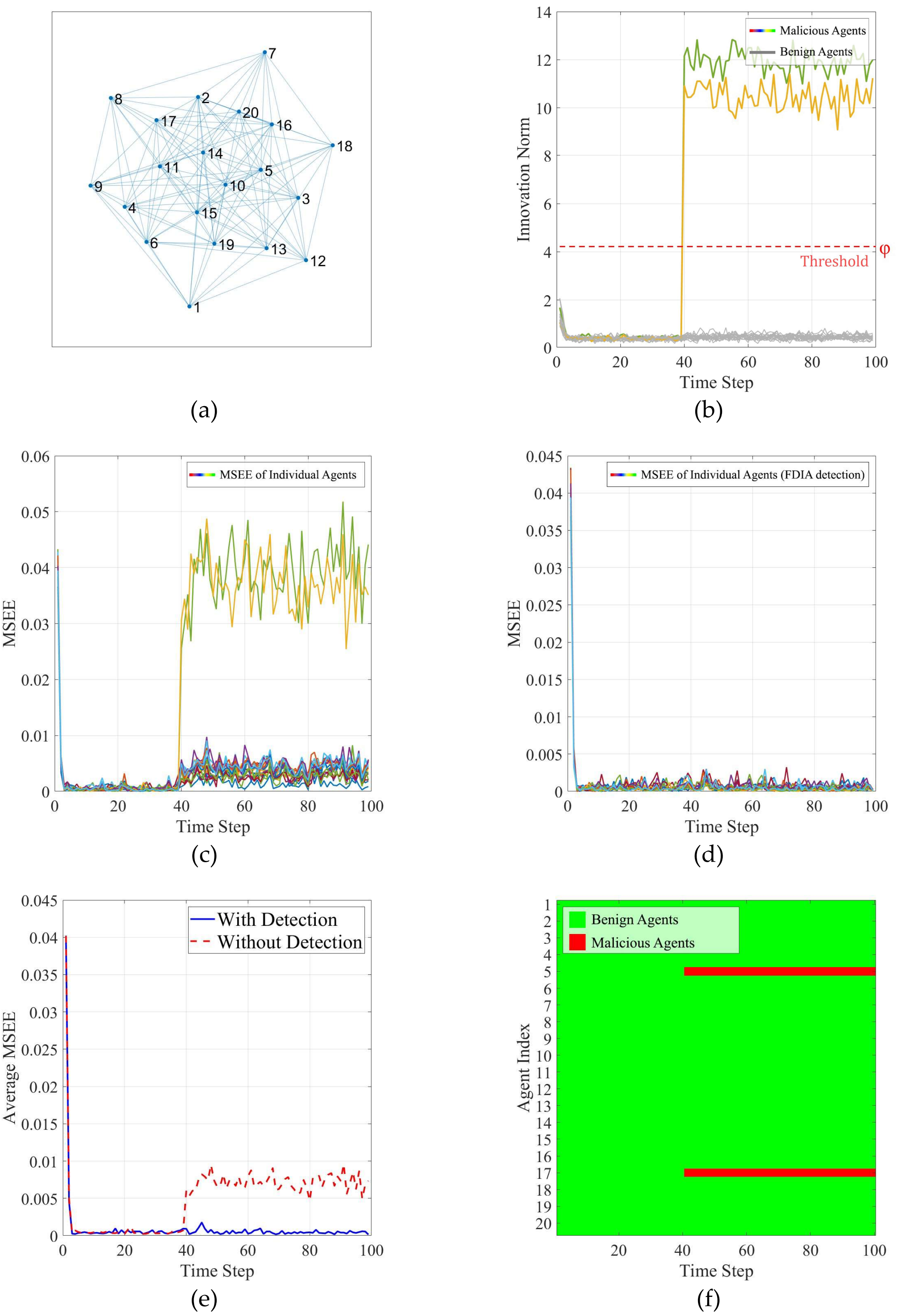

**Figure 5.** Effect of localized false data injection on the proposed algorithms: (a) Random network topology; (b) Normalized innovation values for individual agents compared against the detection threshold $\varphi$, where spikes above the threshold (red dashed line) indicate agents under false data injection attack (colored curves), while values below the threshold (gray curves) correspond to benign agents; (c) Mean squared estimation error (MSEE) over time for individual agents for the baseline algorithm without FDIA detection (shown in colored curves); (d) MSEE over time for individual agents for the proposed algorithm with FDIA detection enabled (shown in colored curves); (e) Comparison of average MSEE of all agents for both algorithms, where the blue line denotes the FDIA-

enabled method and the red dashed line denotes the baseline; (f) Attack detection map ("heat map"), where red regions indicate malicious agents and green regions indicate benign agents.

#### 4.2.2. Case Study 5 – Widespread Coordinated Attacks

In this case study, half of the agents in the communication network (Fig. 6-a) are randomly selected and subjected to false data injection attacks of varying magnitudes (low, medium, and high). The objective is to evaluate the algorithms under severe conditions where a large number of agents are compromised. Particular attention is given to assessing the performance of the proposed detection algorithm when the attack magnitudes approach the detection threshold.

Figure 6-b displays the innovation norm measurements across all agents, showing the presence of attacks with different magnitudes. The MSEE results for both algorithms are shown in Figures 6-c and 6-d, and a comparison of the average MSEE is provided in Fig. 6-e. In the case of low-magnitude attacks, which are close to the detection threshold, some of the compromised agents are successfully detected and isolated. In contrast, others remain undetected and continue to influence the estimates. Even under these challenging conditions, the algorithm with the detection mechanism outperforms the baseline version that lacks detection, achieving significantly lower estimation errors. Finally, Fig. 6-f presents a heat map indicating all detected attacks over time, illustrating the effectiveness of the proposed algorithm in mitigating widespread coordinated false data injections.

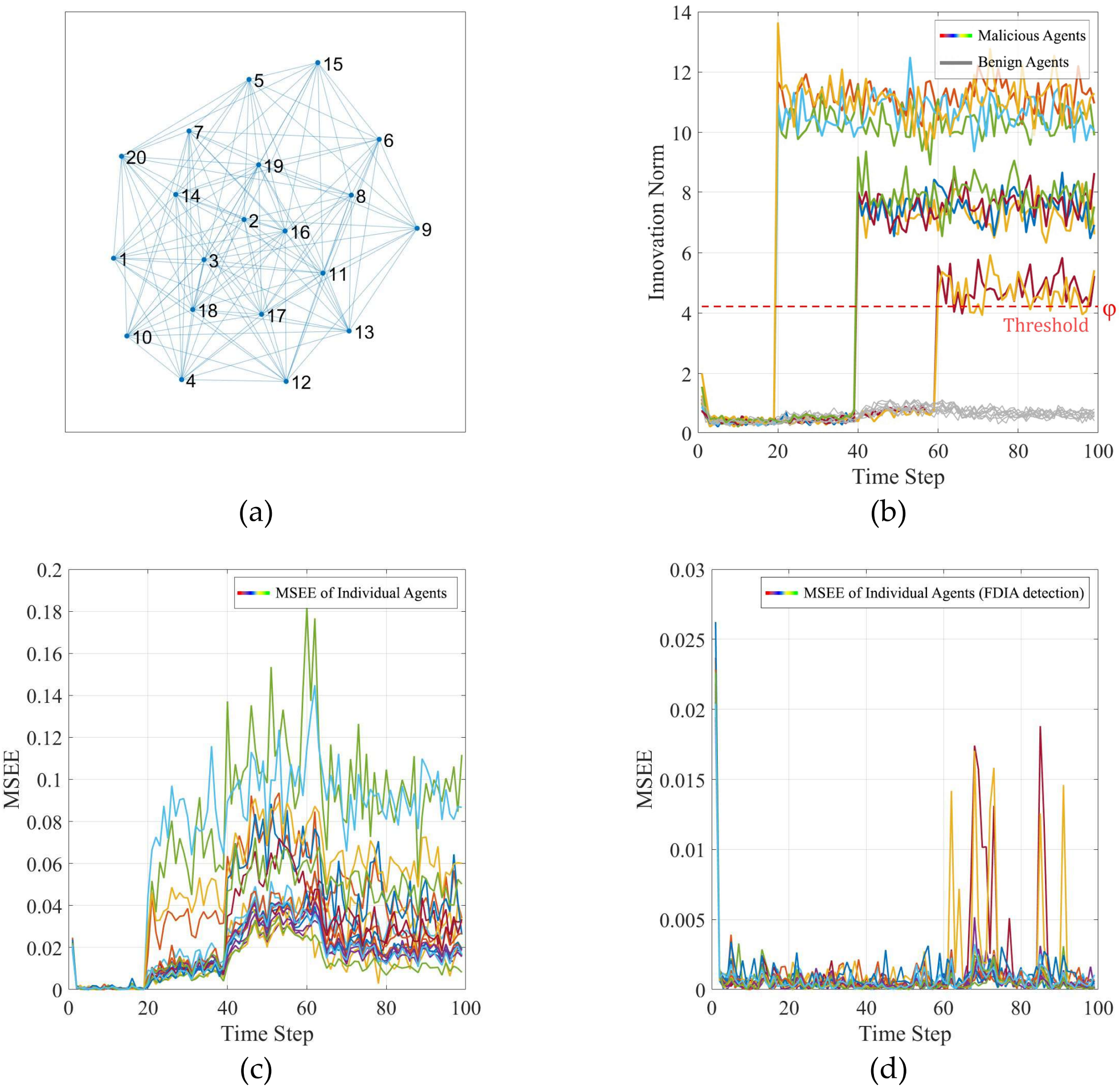

(a) (b)

(c) (d)

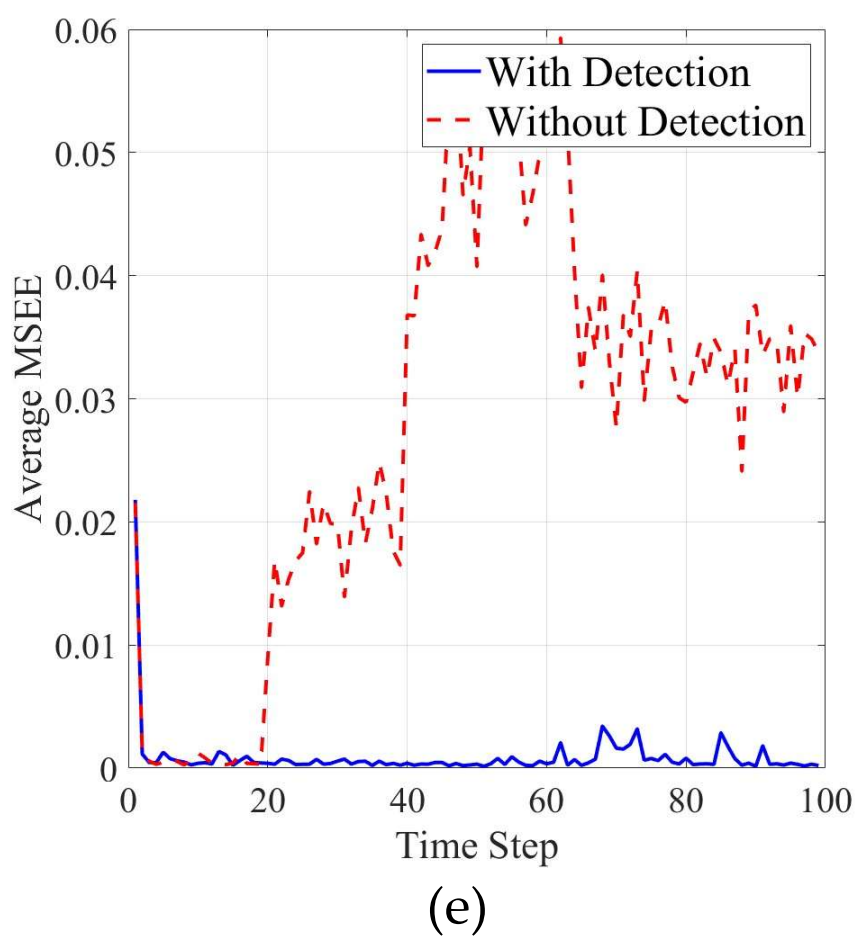

(e)

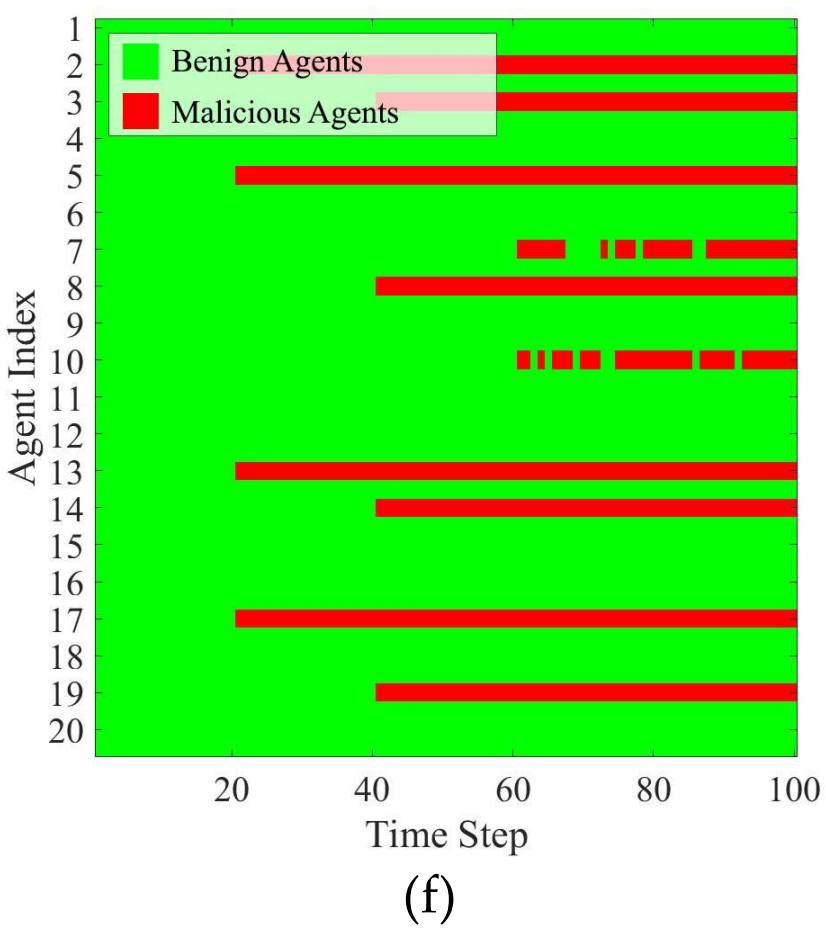

(f)

**Figure 6.** Effect of widespread coordinated attacks on the proposed algorithms: (a) Random network topology; (b) Normalized innovation values for individual agents compared against the detection threshold $\varphi$, where spikes above the threshold (red dashed line) indicate agents under false data injection attack (colored curves), while values below the threshold (gray curves) correspond to benign agents; (c) Mean squared estimation error (MSEE) over time for individual agents for the baseline algorithm without FDIA detection (shown in colored curves); (d) MSEE over time for individual agents for the proposed algorithm with FDIA detection enabled (shown in colored curves); (e) Comparison of average MSEE of all agents for both algorithms, where the blue line denotes the FDIA-enabled method and the red dashed line denotes the baseline; (f) Attack detection map ("heat map"), where red regions indicate malicious agents and green regions indicate benign agents.

#### 4.2.3. Case Study 6 – Transient Attacks

In this final robustness case study, we evaluate the performance of the FDIA detection algorithm under transient attacks. Unlike the previous case studies, where the attacker's influence continuously throughout the entire process, this scenario considers a situation where a subset of agents in the network (Fig. 7-a) experiences short-duration false data injection attacks with medium to high magnitudes. The corresponding innovation norm is shown in Fig. 7-b.

Figures 7-c and 7-d present the MSEE for the algorithms without and with FDIA detection, respectively. A comparison of the average MSEE for both algorithms is displayed in Fig. 8-e. The results demonstrate that the algorithm with detection enabled successfully prevents transient attacks from significantly affecting the estimation process and is able to quickly detect the attacked agents, as indicated in Fig. 7-f.

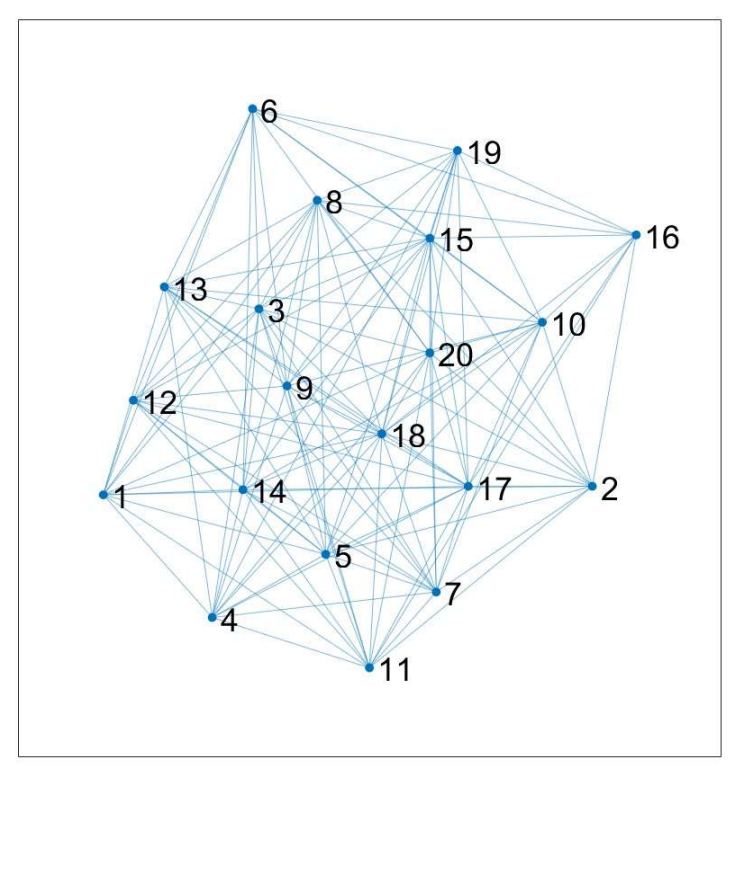

(a)

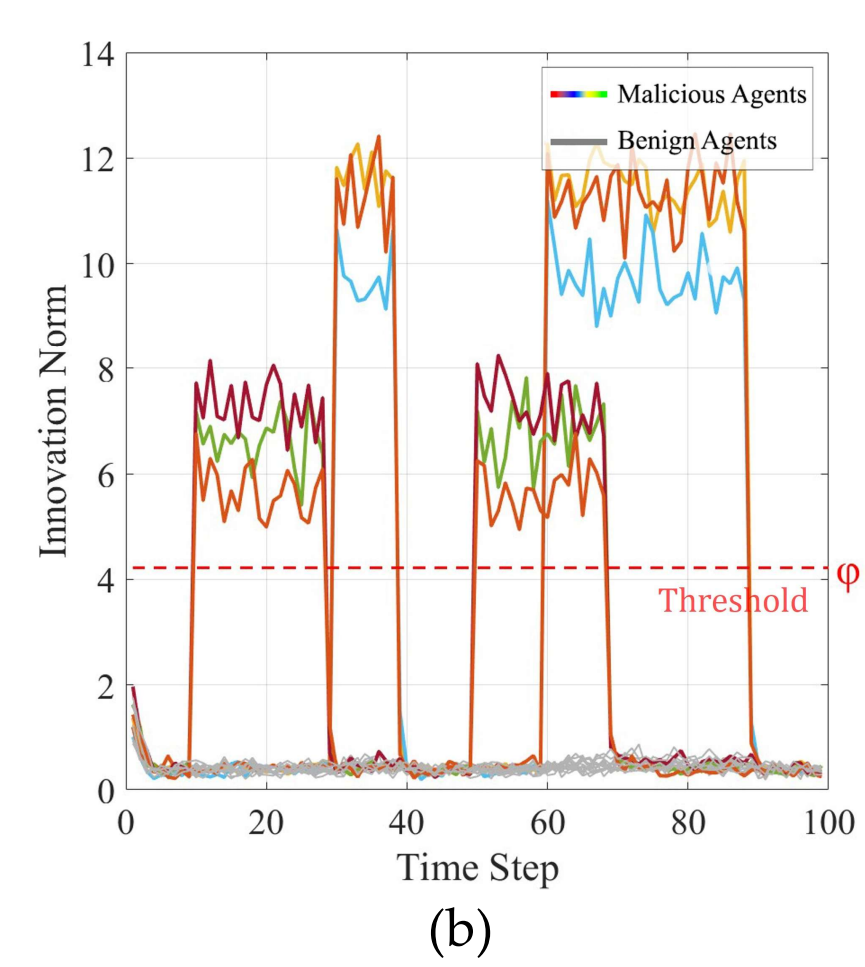

(b)

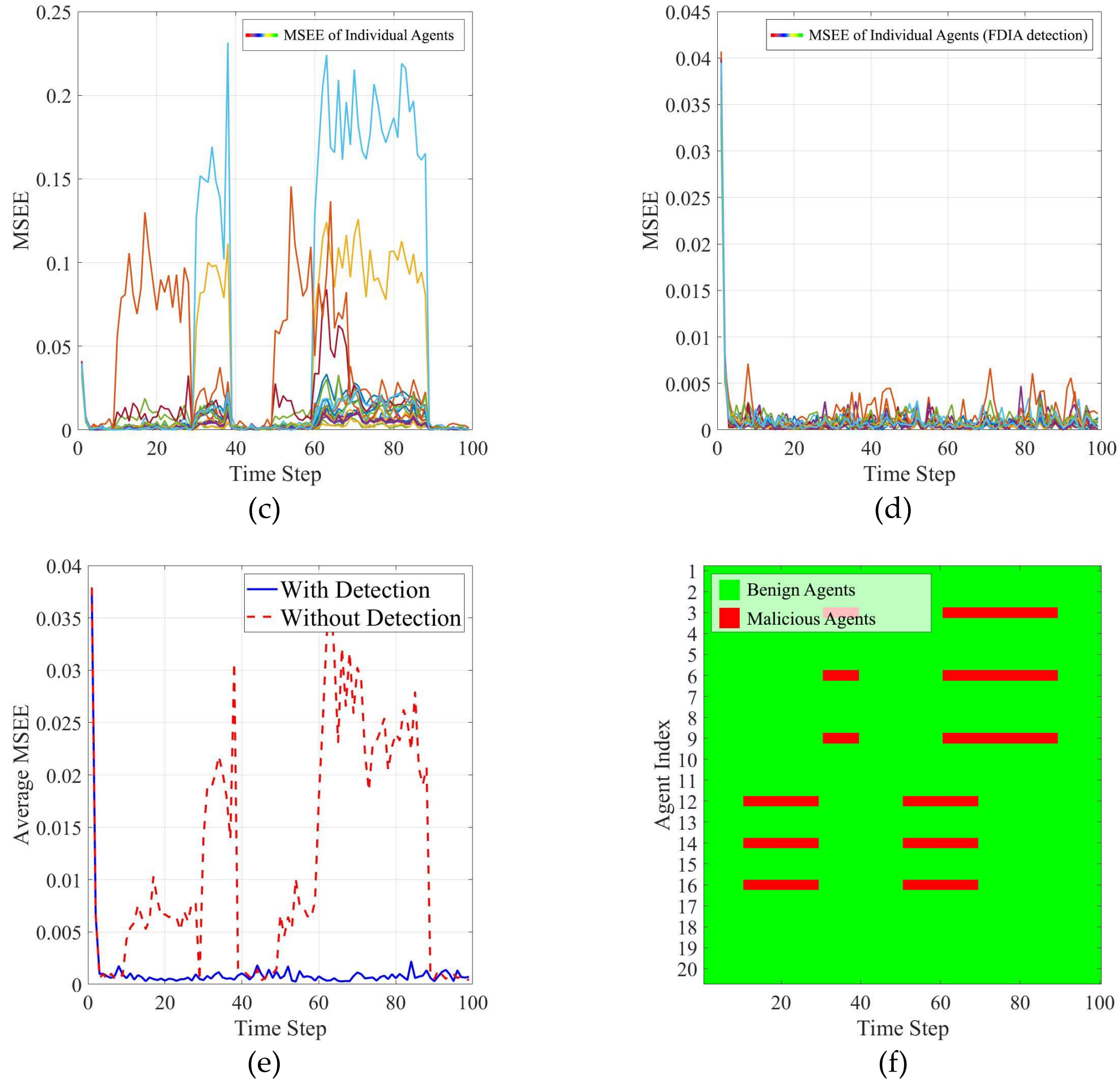

**Figure 7.** Effect of transient attacks on the proposed algorithms: (a) Random network topology; (b) Normalized innovation values for individual agents compared against the detection threshold $\varphi$, where spikes above the threshold (red dashed line) indicate agents under false data injection attack (colored curves), while values below the threshold (gray curves) correspond to benign agents; (c) Mean squared estimation error (MSEE) over time for individual agents for the baseline algorithm without FDIA detection (shown in colored curves); (d) MSEE over time for individual agents for the proposed algorithm with FDIA detection enabled (shown in colored curves); (e) Comparison of average MSEE of all agents for both algorithms, where the blue line denotes the FDIA-enabled method and the red dashed line denotes the baseline; (f) Attack detection map ("heat map"), where red regions indicate malicious agents and green regions indicate benign agents.

### *4.3. Summary and Discussion of Results*

This section summarizes the key findings from the baseline and robustness case studies presented earlier. The baseline case studies assessed the performance of Algorithm (1) under attack-free conditions. In the first baseline case, increased network connectivity led to faster convergence and lower mean squared state estimation error, although this also increased communication overhead. In the second baseline case, the impact of the number of consensus iterations was examined. The results showed that increasing the number of iterations accelerated agreement among agents and reduced estimation errors, but required greater communication bandwidth and power consumption. In the third baseline case, the influence of the saturation filter parameters was investigated. A more conservative threshold achieved higher resilience to measurement variations but also reduced responsiveness to legitimate changes.

The second set of case studies focused on the performance of Algorithm (1) alongside its enhanced version with FDIA detection presented in Algorithm (2), under cyber-physical threats. The fourth case introduced localized high-magnitude false data injection attacks on two agents starting at time step $k > 40$. Algorithm (2) successfully detected and isolated the compromised agents and maintained accurate tracking estimates. The fifth case involved widespread coordinated attacks on 50% of the agents with varying attack magnitudes. The results showed that the medium and high-magnitude attacks were reliably detected and mitigated, while a few low-magnitude attacks near the detection threshold occasionally evaded detection. Even under these challenging conditions, Algorithm (2) achieved substantially better average MSEE and tracking accuracy than Algorithm (1). Finally, the sixth case investigated the algorithms performance under transient attacks. The enhanced algorithm responded quickly to these brief anomalies and preserved estimation performance with minimal delay.

Table 2 concisely summarizes the different scenarios investigated in this study, including both baseline and adversarial conditions explored through simulation. This table highlights the key experimental setups, such as graph connectivity, consensus steps, and attack types, along with the primary outcomes and performance observations for each case. It is important to note that all scenarios are evaluated under the assumption that a majority of agents in the network are benign, which is a standard requirement for consensus-based resilience.

**Table 2.** Summary of Baseline and Robustness Case Studies.

| Scenario | FDA Used | Attack Type | Key Parameters | Main Outcome |
|---|---|---|---|---|
| 1. Graph Connectivity | No | None | p=0.3 vs p=0.7 | Dense graphs improved convergence and lowered MSEE |
| 2. Consensus Iterations | No | None | L=10 vs L=100 | More iterations improved accuracy at increased communication cost |
| 3. Saturation Filter Parameters | No | None | $\xi = 4$ vs $\xi = 2$ | Conservative filters enhanced robustness but reduced responsiveness |
| 4. Localized Fault Injection | Yes | Targeted high-magnitude FDI on 2 agents | Fault at k = 30 | Algorithm (2) detected and suppressed the faulty data |
| 5. Widespread Attack | Yes | Randomized FDI (mixed magnitude) on 50% of agents | FDI levels varied (low/med/high) | Algorithm (2) detected most attacks; low-magnitude injections sometimes evaded detection but accuracy was preserved |
| 6. Transient Attack | Yes | Random short-duration FDI | Limited attack duration | Algorithm (2) suppressed transient anomalies with minimal delay |

## 5. Discussion

While the proposed algorithm demonstrates strong resilience against various false data injection attack scenarios, several limitations should be acknowledged to provide a balanced assessment. First, the method assumes that a majority of agents in the network are benign, a common condition for consensus-based resilience, which may not hold in highly compromised environments. Second, although computational and communication overheads remain modest for small- to medium-scale networks, scaling to very large networks may require additional optimization to prevent excessive processing and bandwidth usage. Third, the approach relies on a Nearly-Constant-Velocity (NCV) motion model, which can reduce estimation accuracy for targets exhibiting highly nonlinear or rapidly changing dynamics; replacing or augmenting the NCV model with a Nearly-Constant-Acceleration (NCA) model or adaptive/learning-based models could improve tracking performance in such scenarios. Finally, the detection mechanism's performance may degrade when attack magnitudes are close to the detection threshold, occasionally resulting in missed detections, as observed in the widespread attack case study. Addressing these limitations could guide future research toward enhancing robustness, scalability, and adaptability of decentralized target tracking algorithms under adversarial conditions.

## 6. Conclusions

In This paper, we developed and analyzed a resilient decentralized target tracking algorithm that operates effectively under both benign and adversarial conditions. By combining a saturation-based consensus filter with a false data injection detection and isolation mechanism, the proposed algorithm can achieve accurate estimates across multi-agent networks even in the presence of localized, coordinated, and transient false data injection attacks. The results from the case studies showed that increased graph connectivity and more consensus iterations improve estimation accuracy, while carefully tuned saturation filters balance estimation precision and sensitivity towards faulty measurements. Under cyber-physical attacks, the dynamic detection threshold successfully identified and isolated compromised agents, allowing the system to maintain stable tracking performance and significantly reduce the mean squared estimation errors.

**Author Contributions:** Conceptualization, A.A.G. and M.D.; methodology, A.A.G. and M.D.; software, A.A.G.; validation, A.A.G. and M.D.; formal analysis, A.A.G. and M.D.; investigation, A.A.G. and M.D.; resources, M.D.; data curation, A.A.G.; writing—original draft preparation, A.A.G.; writing—review and editing, A.A.G. and M.D.; visualization, A.A.G.; supervision, M.D.; project administration, M.D. All authors have read and agreed to the published version of the manuscript.

**Funding:** This research received no external funding.

**Data Availability Statement:** No new data were created or analyzed in this study. The MATLAB simulation code and relevant configuration files used to generate the results are available from the authors upon request.

**Conflicts of Interest:** The authors declare no conflicts of interest.